\documentclass{article}
\usepackage{spconf,amsmath,graphicx}
\usepackage{algorithm}
\usepackage{algpseudocode}
\usepackage{booktabs}
\usepackage{amsfonts}
\usepackage{hyperref}

\title{Large-Scale MIDI-based Composer Classification}
\name{Qiuqiang Kong, Keunwoo Choi, Yuxuan Wang}
\address{ByteDance, Mountain View, California, USA \\ \{kongqiuqiang, keunwoo.choi, wangyuxuan.11\}@bytedance.com}

\begin{document}
\maketitle
%
% \ninept

\begin{abstract}
Music classification is a task to classify a music piece into labels such as genres or composers. We propose large-scale MIDI based composer classification systems using GiantMIDI-Piano, a transcription-based dataset. We propose to use piano rolls, onset rolls, and velocity rolls as input representations and use deep neural networks as classifiers. To our knowledge, we are the first to investigate the composer classification problem with up to 100 composers. By using convolutional recurrent neural networks as models, our MIDI based composer classification system achieves a 10-composer and a 100-composer classification accuracies of 0.648 and 0.385 (evaluated on 30-second clips) and 0.739 and 0.489 (evaluated on music pieces), respectively. Our MIDI based composer system outperforms several audio-based baseline classification systems, indicating the effectiveness of using compact MIDI representations for composer classification.

\end{abstract}
\section{Introduction}
\label{sec:intro}
Music classification is a task to classify music pieces into a certain set of categories such as genres \cite{tzanetakis2002musical}, composers, mood, or tags \cite{lamere2008social}. As a tool for music listeners and creators, music classification can be deployed as a part of music recommendation systems. With the overwhelming sizes of available online music catalogues, music classification has been gaining a vast amount of attention in music research community and industry \cite{nam2018deep}.  %  many applications in recommending systems, data mining, music information retrievals, etc. Music classification has been widely researched in recent years, and has been applied to several products in the mobile. 

A majority of music classification research has been performed on the audio signal domain. This may has been a practical choice because the audio signal is the final form of produced music for most of music genres. 
% An exception would be classical music, which is usually published first as a form of symbolic representation -- the score. Even so, practically, in music consumption, classical music is distributed in the form of audio signal. 
% Early audio tagging systems were developed based on audio features such as MFCCs and spectral features \cite{tzanetakis2002musical}. 
For example, deep learning based methods have been proposed and have achieved state-of-the-art performances in various music classification tasks \cite{dieleman2014end, choi2016automatic, lee2017sample}. Similarly, many datasets with audio signals and corresponding labels have been created such as GTZAN \cite{tzanetakis2002musical} and FMA \cite{defferrard2016fma} for genre classification, IRMAS \cite{burred2009dynamic} and OpenMIC \cite{humphrey2018openmic} for instrument recognition, and MSD \cite{bertin2011million} and MagnaTagATune \cite{law2009evaluation} for tagging.

In this paper, although it has been extremely popular, we hypothesize that using audio signals might introduce an inherent limitation to build composer classification systems. Audio signal is rather an entangled representation of music. One may see music as a combination of a symbolic representation and a timbre. Symbolic representations usually include musical notes and notations and are used to be the final form of music production in classical music. This means that a composer classification system can benefit from symbolic representations by focusing on an aspect of music that is the most relevant to the composer. In addition, the compactness of symbolic representations may lead to a more efficient system in terms of computation and memory usage.

In practice, the progress of music classification using symbolic representations has not been very fast. This is partly due to the lack of datasets. There exist a few datasets that include symbolic representations of classical music \cite{hawthorne2018enabling, emiya2010maps}, but the number of composers of these datasets are limited. For example, a recently released the MAESTRO dataset includes 61 classical composers \cite{hawthorne2018enabling}. As a result, the existing works have shown some limited scalabilities \cite{cataltepe2007music, herremans2016composer, kim2020deep} such as the small number of unique music pieces. There are two issues with the small scale experiments. First, it is difficult to expect that the trained models would generalize well with unseen examples, i.e., the models can overfit to the whole dataset. Second, the wisdom such as optimization method or architectural choice might not be applicable to large scale experiments in the real world.

% Music classification has been a popular research topic in recent years. Early work of music classification apply time-frequency based methods, such as manuualy selected features. Then, classifers are built on the individual frames. Recently, neural network based methods have been applied for music classification, such as fully connected neural networks, recurrent neural networks and convolutional neural networks. Those neural network based methods have achieved state-of-the-art performance in several audio tagging tasks. Usually, log mel spectrograms are used features to be fed into those neural networks. 

% Most of music classification datasets are based on audio recordings, such as the GTZAN dataset, the Million Song Dataset (MSD), the xxx dataset. Those datasets either provide audio recordings or release the features for users to build classification systems. However, there is a lack of research on investigating symbolic datasets such as MIDI datasets for music classification. the symbols from composers convey rich information of who a composer is. One reason of the lack of research is that there is a lack of datasets containing such a large number of MIDI files to build the system. Previous MIDI base composer classification systems are mostly focused on relative small datasets.

In this article, we propose large-scale and MIDI-based composer classification systems for 10 and 100 composers. The systems are trained and evaluated on GiantMIDI-Piano dataset \cite{kong2020giantmidi}, the largest public classical piano MIDI dataset to date. 
The novelty of this work is that we compare audio based log mel spectrogram features with transcribed MIDI representations for composer classification. 
We propose to use frame rolls, onset rolls, and velocity rolls extracted from MIDI files as input feature. We report that our systems with MIDI input outperform the baseline systems with the log mel spectrogram feature. We show that the classification result of different composers varies. % Finally, we visualize the embedding vectors of MIDI clips composed by different composers.

This paper is organized as follows. Section 2 introduces the dataset and input representations. Section 3 introduces the details of the deep neural network architectures we used. Section 4 shows experimental results and Section 5 concludes the work. 

\section{Dataset and representation}
\subsection{GiantMIDI-Piano dataset}
Previous composer classification systems \cite{cataltepe2007music, herremans2016composer, kim2020deep} were developed on relative small datasets which only include tens of composers.
%and X items. 
In other words, There is a lack of research on large-scale composer classification. Recently, 
% we 
Kong et al. have released a large-scale MIDI dataset called  GiantMIDI-Piano \cite{kong2020giantmidi}. GiantMIDI-Piano is collected as follows. 1) The names of pieces and composers are collected from the International Music Score Library Project (IMSLP). 2) Audio recordings are downloaded from YouTube by searching the names of pieces and composers. 3) A piano solo detection system is applied to detect piano solos. 4) Piano solos are transcribed to MIDI files using a high-resolution piano transcription system \cite{kong2020high}. In total, there are 10,854 transcribed MIDI files composed by 2,786 composers. The duration of music pieces and number of composers of GiantMIDI-Piano are significantly longer and larger than previous piano MIDI datasets \cite{hawthorne2018enabling, emiya2010maps}.

\subsection{Data representation}
\begin{figure}[t]
  \centering
  \centerline{\includegraphics[width=\columnwidth]{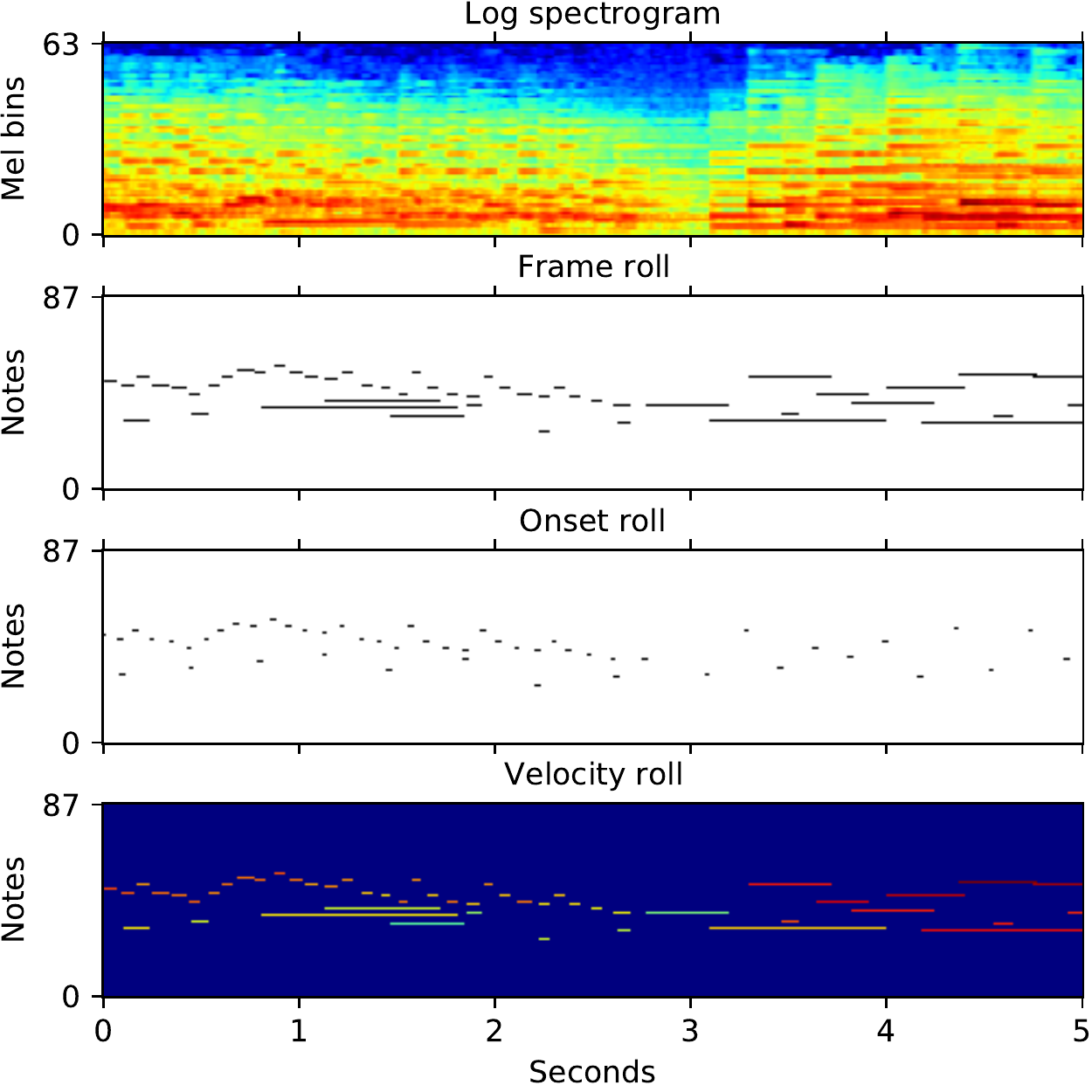}}
  \caption{From top to bottom -- A log mel spectrogram, a MIDI frame roll, a MIDI onset roll, and a MIDI velocity roll of a 5-second audio clip in GiantMIDI-Piano dataset.}
  \label{fig:rolls}
\end{figure}

For audio based composer classification, we use log-magnitude mel spectrograms as input feature following previous research \cite{choi2018comparison, choi2016automatic, kong2019panns, dieleman2014end}. We denote a log mel spectrogram as $ X_{\text{mel}} $ with a shape of $ T \times F $, where $ T $ is the number of frames and $ F $ is the number of mel bins. The first row of Fig.~\ref{fig:rolls} shows the log mel spectrogram of a 5-second audio clip.

For MIDI based composer classification, 
% the representations are different. MIDI files contains lists of music events. Each event is represented by a tuple. For example, a note event can be represented by a tuple of $<$pitch, velocity, time\_stamp$>$. Several previous works have applied these sparse representations for expressive pianists modeling \cite{oore2020time, jeong2019virtuosonet}.
% In this work, 
we follow and extend previous works \cite{walder2016modelling, kim2020deep} using piano rolls  as input features.
% Piano rolls are continuous rolls of paper with perforations punched into them, and were developed to record music that can be played on a musical instrument. 
A piano roll is in a shape of $ T \times K $, where $ T $ is the number of temporal frames and $ K $ is number of pitches which is set to 88, the number notes on a piano. Based on what it represents, there are so called a frame roll, an onset roll, and a velocity roll. 
% A piano roll has a time dimension and a pitch dimension. 
In this work, we 
% propose to 
use all three for them as well as their combinations as input representations to the proposed systems.

\begin{figure*}[t]
  \centering
  \centerline{\includegraphics[width=\textwidth]{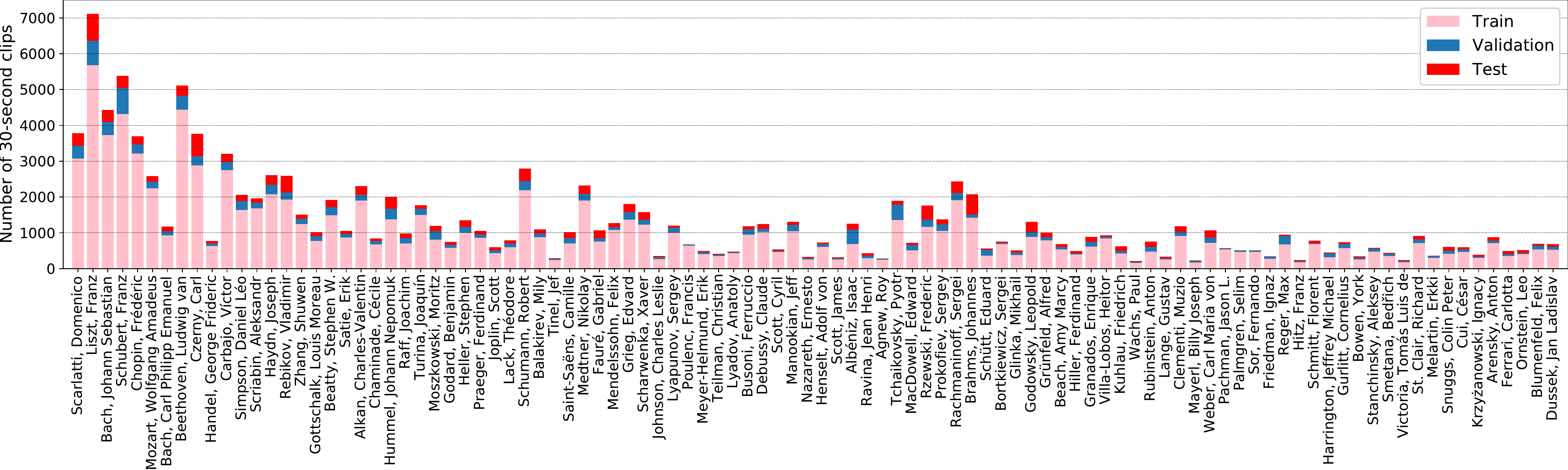}}
  \caption{Number of training, validation and testing 30-seconds of top 100 composers sorted by their number of works in descending order.}
  \label{fig:clips_num}
\end{figure*}

% \subsubsection{Frame roll}
A \textbf{frame roll}, $ X_{\text{frame}} $, is binarized to $1$ or $0$ and represents the activation of piano notes in each temporal frame. 
% The elements of $ X_{\text{frame}} $ have values of 1 or 0, indicating the active or inactive of piano notes in the each frame.
%
% \subsubsection{Onset roll}
% In addition to frame rolls, 
An \textbf{onset roll}, $ X_{\text{onset}} $, provides onset information of every played note and is binarized as well. Previous works have shown that onsets are important for music understanding such as piano transcription \cite{hawthorne2018enabling, kong2020high} and we expect onset rolls would provide supplementary information to frame rolls.
% We denote an onset roll with a shape of $ T \times K $ as . The elements of $ X_{\text{onset}} $ have values of 1 or 0, indicating the presence or absence of onsets in the onset roll. 
%
% \subsubsection{Velocity roll}
Finally, a \textbf{velocity roll}, $ X_{\text{vel}} $, represents the recorded velocity in the MIDI format. The original value in MIDI recording ranges from 0 to 127 but we normalized it to $[0, 1]$.
% In addition to frame rolls and onset rolls, 
Articulations are important parts of music pieces that can be represented in velocity rolls. For example, composers may have distinctive patterns in using dynamics in their works. 
% Velocity provides rich information of music pieces composed by different composers. We denoe 
% a velocity roll with shape of $ T \times K $ as $ X_{\text{vel}} $. In the MIDI format, velocities are represented with integers between 0 and 128. We normalize those velocities to continuous values between 0 and 1. The elements of $ X_{\text{vel}} $ have values between 0 and 1, indicating dynamics from pianissimissimo to fortissimissimo. 
The second, third, and fourth rows of Fig. \ref{fig:rolls} illustrates the frame, onset, and velocity roll of a 5-second MIDI clip, respectively.

% \subsubsection{Concatenated rolls}
In the experiment, the final input representation of an item is a multi-channel $T \times K$ image where each roll corresponds to each channel.

\begin{table*}[t]
\centering
\caption{Class-wise accuracies of the 10-composer classification systems (CRNN)}
\label{table:classwise_results}
\resizebox{\textwidth}{!}{\begin{tabular}{l*{11}{c}}
 \toprule
 & Scarlatti & Liszt & J. S. Bach. & Schubert & Chopin & Mozart & C. P. E. Bach & Beethoven & Czerny & Handel & Avg. \\
 \midrule
 Audio (Log mel spectrogram) & \textbf{0.886} & 0.780 & 0.784 & 0.460 & 0.395 & 0.516 & \textbf{0.735} & \textbf{0.384} & 0.640 & \textbf{0.618} & 0.620 \\
 MIDI (frame) & 0.834 & 0.716 & 0.852 & 0.416 & \textbf{0.682} & 0.497 & 0.621 & 0.329 & 0.683 & 0.559 & 0.619 \\
 MIDI (onset) & 0.834 & 0.770 & \textbf{0.864} & 0.378 & 0.558 & 0.484 & 0.583 & 0.353 & 0.643 & 0.544 & 0.601 \\
 MIDI (frame + onset) & 0.874 & 0.764 & 0.802 & 0.392 & 0.592 & 0.497 & 0.636 & 0.318 & 0.711 & 0.544 & 0.613 \\
 MIDI (frame + onset + velocity) & 0.829 & \textbf{0.796} & 0.831 & \textbf{0.555} & 0.597 & \textbf{0.584} & 0.705 & 0.240 & \textbf{0.759} & 0.588 & \textbf{0.648} \\
 \bottomrule
\end{tabular}}
\end{table*}

\section{Composer classification model}\label{sec:model}
We use the same Convolutional Neural Networks (CNNs) architecture for both audio-based and MIDI-based classification. 
The only difference is the number of input channel, which is 1 for audio based classification and equal to the number of rolls used (3 in the reference system which uses all the rolls) in MIDI based classification.
% \subsection{Audio based composer classification}
% We use CNNs as classifiers for audio based composer classification using log mel spectrogram as features. 

% We denote the target of an audio clip as $ y \in \{0, 1\}^{K} $, and the model output as $ f(X_{\text{mel}}) \in [0, 1]^{K} $.
The CNN consists of 8 convolutional layers \cite{choi2016automatic, kong2019panns}, each of which consists of a linear convolution operation with a kernel size of $ 3 \times 3 $, a ReLU nonlinearity \cite{nair2010rectified}, and a batch normalization \cite{ioffe2015batch}.
% to speed up and stablize training. 
Average pooling layers with shapes of $ 2 \times 2 $ are applied after every two convolutional layers. A global max pooling operation is applied after the last convolutional layer to summarize feature maps to an embedding vector. Then, two fully connected layers followed by a softmax non-linearity are used to predict the output probabilities of composers. 
% In training, we apply a categorical crossentropy to train the systems.
We set the numbers of output channels of convolutional layers to 64, 64, 128, 128, 256, 256, 512, and 512. Dropout \cite{srivastava2014dropout} rates of 0.2 and 0.5 are used after all the convolutional blocks and fully connected layers, respectively.

Vanilla CNNs usually have limited receptive field size and can not capture a long temporal dependency of audio signals. To address this problem, we propose to apply recurrent neural networks after convolutional layers to increase the representation abilities of models which are called convolutional recurrent neural networks (CRNNs) \cite{choi2017convolutional}. Convolutional layers are used for extracting high-level features from inputs and recurrent layers are used for capturing long time dependencies of log mel spectrogram or MIDI rolls. We use bidirectional gated recurrent unit (biGRU) to model the recurrent neural network to utilize both past and future information \cite{chung2014empirical}.

% \subsection{Clip-wise and Piece-wise Prediction}
During inference, we input 30-second audio clips or MIDI clips into trained models to predict the presence probabilities of composers on 30-second clips. However, a music piece usually consists of several 30-second clips. We aggregate the clip-wise predictions by directly averaging the probability outputs of a model over the clips.

% We propose an aggregating algorithm to ensemble clip-wise predictions to piece-wise predictions. To begin with, we split a music piece into 30-second clips. We denote the predictions of clips as $ \{ p_{1}, p_{2}, ..., p_{I} \} $, where $ p_{i} \in [0, 1]^{K} $ is the predicted probability of $ K $ composers, and $ I $ is the number of 30-second clips in the music piece. Then, the prediction of music pieces can be obtained by:

% \begin{equation} \label{eq:aggregate}
% q = \frac{1}{|I|}\sum_{i=1}^{I} p_{i},
% \end{equation}

% \noindent where $ q \in [0, 1]^{K} $. In experiments, we show and analyses both clip-wise and piece-wise composer classification results.

% \begin{equation} \label{eq1}
% l = -\sum_{k=1}^{K}y_{k} \text{ln} f(X_{\text{mel}})
% \end{equation}

% \subsection{MIDI based composer classification}
% We use CNNs as classifiers for MIDI based composer classification using concatenated rolls described in (\ref{eq:concat_roll}) as input features. The architecture of the MIDI based composer classification system is the same as the audio based composer classification system, except that there are three input channels to the CNN, corresponding to the frame rolls, onset rolls and velocity rolls respectively. By using the same CNN architecture, we could fairly compare using log mel spectrogram and MIDI representations as inputs for composer classification.

\section{Experiments}
\subsection{Overview}
We perform experiments to test the proposed MIDI based composer classification systems on the GiantMIDI-Piano dataset \cite{kong2020giantmidi}.
% There are 2,786 composers in total in GiantMIDI-Piano. 
% The number of piano pieces composed by different composers have a long tail distribution. 
For the 10-composer and 100-composer classification systems, out of 2,786 composers in the dataset, we select top 10 and 100 composers based on the numbers of piano pieces. The 100 composers are listed at the horizontal axis of Fig.~\ref{fig:clips_num}.
% build 10-composer and 100-composer classification systems for the composers who have the most number of piano pieces in GiantMIDI-Piano.
The dataset is split into training, validation, and test subsets with a ratio of 8:1:1 in a stratified fashion.
We segment every music piece into 30-second clips for training and testing. Each clip inherits the composer label of its parent music piece. Fig.~\ref{fig:clips_num} shows the numbers of 30-second clips in training, validation and test set for the top-100 composers. 
% Some composers such as Liszt has over 5000 clips for training. 
In total, there are 31,127 30-second clips for training 10-composer classification systems and 101,417 30-second clips for training 100-composer classification systems.

% Audio clips are downmixed to monophonic and resampled. to 16 kHz that is consistent with the piano transcription procedure  \cite{kong2020high} to transcribe the GiantMIDI-Piano dataset. 
For the audio based system, we compute the magnitude of a short time Fourier transform (STFT) with a Hann window size 1024 and a hop size of 160. Then, we apply 64 mel filter banks on the spectrograms followed by a logarithm operation to extract log mel spectrograms.
To obtain MIDI representation $ X_{\text{roll}} $, we parse the onset, offset and velocity information of notes from transcribed MIDI files to obtain $ X_{\text{frame}} $, $ X_{\text{onset}} $, and $ X_{\text{vel}} $ to constitute $ X_{\text{roll}} $. 

During training, we use a categorical crossentropy loss, an Adam optimizer \cite{kingma2014adam} with a learning rate of 0.001, and a batch size of 16.
% All systems are trained for 200,000 iterations.

\subsection{Results}

Table \ref{table:classwise_results} shows the class-wise classification accuracies of 10-composer classification system. All systems are based on CRNN architectures with different inputs. The classification results noticeably vary among systems and composers. We observe two patterns from this result. First, some composers are more easily distinguishable than others. For example, the classification accuracy of Scarlatti are over 0.82 for all systems. This was also observed in \cite{kim2020deep}.
Second, for the classification of each composer, the importance of frame, onset, and velocity varies, and furthermore, some of the information even hurt the performance. For example, the classification accuracies of Liszt, Schubert, Mozart, and Czerny are higher when all the MIDI rolls are used than other MIDI-based systems, indicating that velocity plays an important roll in classifying Liszt. This also holds for the average accuracy. However, for J.~S.~Bach, Chopin, Beethoven, C.~P.~E.~Bach, and Handel, adding the velocity roll to the frame + onset system degraded the performance. 

\begin{figure}[t]
  \centering
  \centerline{\includegraphics[width=\columnwidth]{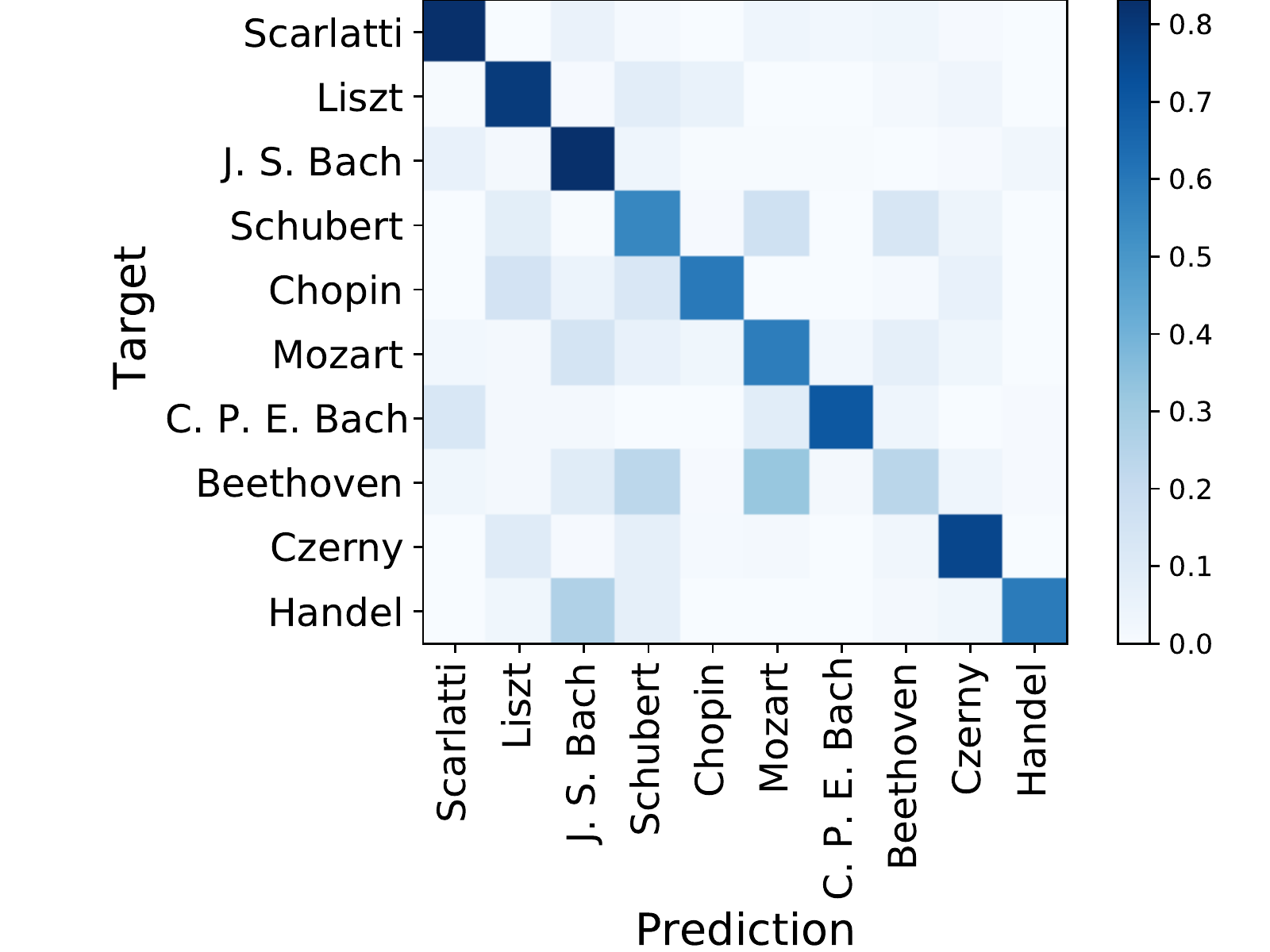}}
  \caption{Confusion matrix of 10-composer classification with CRNN using MIDI frame + onset + velocity rolls as input.}
  \label{fig:confusion_matrix}
\end{figure}

Fig. \ref{fig:confusion_matrix} shows the confusion matrix of 10-composer classification result. Beethoven is often confused with Schubert and Mozart. However, the model is less misclassify the pieces by Schubert and Mozart as Beethoven. This might be due to the exceptionally wide range of musical style Beethoven's pieces have, but a deeper analysis would be necessary to confirm this conjecture. The pieces of Handel turns out to be mostly confused with the pieces of J. S. Bach, who are both from Baroque era. A similar pattern has been also observed in \cite{kim2020deep}.

Table \ref{table:clips_results} shows the `macro'-averaged classification accuracies of 30-second clips of all composers. Macro-average is obtained by first calculating the classification accuracies of individual composers, then, averaging the them over all the composers. Table \ref{table:clips_results} shows that the CRNN performs better than the CNN in both 10-composer and 100-composer classification tasks. The CNN and CRNN systems with log mel spectrogram achieve 10-composer classification accuracies of 0.589 and 0.620, respectively. With using frame + onset + velocity MIDI rolls, the systems achieve accuracies of 0.604 and 0.648, respectively. On the other hand, log mel spectrogram outperforms MIDI representations as input for 100-composer classification. One reason might be because there are several music pieces played by similar pianos in the training and testing set, which can affect more in 100-composer classification due to its wide coverage of the dataset. Log mel spectrogram-based system learns piano timbre information and achieves better classification result than transcribed piano rolls.

Table \ref{table:piece_results} shows the macro-averaged classification accuracies of music pieces of all composers. The piece-wise predictions are obtained as explained in Section~\ref{sec:model}. For 10-composer classification, the CRNN system using frame + onset + velocity as input achieves the highest accuracy of 0.739. For 100-composer classification, the CRNN system using frame + onset as input achieves the highest accuracy of 0.489. The piece-wise results are higher than 30-second clip-wise results, indicating that by averaging the predictions of 30-second clips, the noise of the prediction introduced by selecting an arbitrary segmentation can be successfully suppressed with multiple segments.

\begin{table}[t]
  \vspace{-6pt}
  \caption{The accuracies of 30-second clip-wise classification}
  \label{table:clips_results}
  \centering
  \resizebox{\columnwidth}{!}{\begin{tabular}{lcccc}
    \toprule
    & \multicolumn{2}{c}{\textbf{\textsc{10-composer}}} & \multicolumn{2}{c}{\textbf{\textsc{100-composer}}} \\
	\cmidrule(lr){2-3} \cmidrule(lr){4-5} 
    & CNN & CRNN & CNN & CRNN \\
    \midrule
 Audio (Log mel spectrogram) & 0.589 & 0.620 & \textbf{0.341} & \textbf{0.385} \\
 MIDI (frame) & 0.573 & 0.619 & 0.303 & 0.322 \\
 MIDI (onset) & 0.571 & 0.601 & 0.281 & 0.290 \\
 MIDI (frame + onset) & 0.598 & 0.613 & 0.317 & 0.338 \\
 MIDI (frame + onset + velocity) & \textbf{0.604} & \textbf{0.648} & 0.324 & 0.342 \\

	\bottomrule
\end{tabular}}
\end{table}

\begin{table}[t]
  \caption{The accuracies of piece-wise composer classification by averaging model outputs}
%   \vspace{6pt}
  \label{table:piece_results}
  \centering
  \resizebox{\columnwidth}{!}{\begin{tabular}{lcccc}
    \toprule
    & \multicolumn{2}{c}{\textbf{\textsc{10-composer}}} & \multicolumn{2}{c}{\textbf{\textsc{100-composer}}} \\
	\cmidrule(lr){2-3} \cmidrule(lr){4-5} 
    & CNN & CRNN & CNN & CRNN \\
    \midrule
 Audio (Log mel spectrogram) & 0.592 & 0.643 & 0.402 & 0.411 \\
 MIDI (frame) & 0.666 & 0.737 & 0.452 & 0.484 \\
 MIDI (onset) & \textbf{0.669} & 0.672 & 0.419 & 0.427 \\
 MIDI (frame + onset) & 0.623 & 0.691 & \textbf{0.467} & \textbf{0.489} \\
 MIDI (frame + onset + velocity) & 0.661 & \textbf{0.739} & \textbf{0.467} & 0.477 \\

	\bottomrule
\end{tabular}}
\end{table}

\section{Conclusion}
In this paper, we proposed large-scale MIDI based composer classification systems of classical piano solo works. This is the first work of investigating transcribed MIDI files from audio recordings for composer classification. The MIDI representations include frame roll, onset roll and velocity rolls (or their subsets), and convolutional neural networks and convolutional recurrent neural networks are used as models to build composer classification systems. We show that the classification accuracies of different composers vary among different systems. Our best system achieved accuracies of 0.739 and 0.489 on 10-composer and 100-composer classification respectively. 
A future direction can be a generalization towards other music genres and instruments, which would expand the scope of the proposed system to popular music and make such a system more widely useful.

% The T-SNE visualization of embeddings show that composers of different genres have different distance in the embedding space. 

\small
\bibliographystyle{IEEEbib}
\bibliography{strings,refs}

\end{document}